\documentclass[manuscript]{aastex631}
\newcommand\wpe{\omega_\mathrm{pe}}
\newcommand\lde{\lambda_\mathrm{De}}
\usepackage{graphicx,epsfig,fancyhdr,epsf,txfonts,epstopdf}
\shorttitle{Verification of the standard theory of plasma emission}
\shortauthors{Zhang et al.}
\graphicspath{{./}{figures/}}

\begin{document}

\title{Verification of the standard theory of plasma emission with particle-in-cell simulations}

\correspondingauthor{Yao Chen}
\email{yaochen@sdu.edu.cn}

\author{Zilong Zhang}
\affiliation{Institute of Space Sciences, Shandong University, Shandong, 264209, People's Republic of China.}
\affiliation{Institute of Frontier and Interdisciplinary Science, Shandong University, Qingdao, Shandong, 266237, People's Republic of China.}

\author{Yao Chen}
\affiliation{Institute of Space Sciences, Shandong University, Shandong, 264209, People's Republic of China.}
\affiliation{Institute of Frontier and Interdisciplinary Science, Shandong University, Qingdao, Shandong, 266237, People's Republic of China.}

\author{Sulan Ni}

\affiliation{Institute of Frontier and Interdisciplinary Science, Shandong University, Qingdao, Shandong, 266237, People's Republic of China.}
\affiliation{Institute of Space Sciences, Shandong University, Shandong, 264209, People's Republic of China.}

\author{Chuanyang Li}
\affiliation{Institute of Frontier and Interdisciplinary Science, Shandong University, Qingdao, Shandong, 266237, People's Republic of China.}
\affiliation{Institute of Space Sciences, Shandong University, Shandong, 264209, People's Republic of China.}

\author{Hao Ning}
\affiliation{Institute of Frontier and Interdisciplinary Science, Shandong University, Qingdao, Shandong, 266237, People's Republic of China.}
\affiliation{Institute of Space Sciences, Shandong University, Shandong, 264209, People's Republic of China.}

\author{Yaokun Li}
\affiliation{Institute of Frontier and Interdisciplinary Science, Shandong University, Qingdao, Shandong, 266237, People's Republic of China.}
\affiliation{Institute of Space Sciences, Shandong University, Shandong, 264209, People's Republic of China.}

\author{Xiangliang Kong}
\affiliation{Institute of Space Sciences, Shandong University, Shandong, 264209, People's Republic of China.}
\affiliation{Institute of Frontier and Interdisciplinary Science, Shandong University, Qingdao, Shandong, 266237, People's Republic of China.}

\begin{abstract}
The standard theory of plasma emission is based on kinetic couplings between a single beam of energetic electrons and unmagnetized thermal plasmas, involving multi-step nonlinear wave-particle and wave-wave interactions. The theory has not yet been completely verified with fully-kinetic electromagnetic particle-in-cell (PIC) simulations. Earlier studies, greatly limited by available computational resources, are controversial regarding whether the fundamental emission can be generated according to the standard theory. To resolve the controversy, we conducted PIC simulations with a large domain of simulation and a large number of macroparticles, among the largest ones of similar studies. We found significant fundamental emission if the relative beam density is small enough (say, $\le$ 0.01), in line with earlier study with a much-smaller domain; the relative intensity (normalized by the total initial beam energy) of all modes, except the mode associated with the beam-electromagnetic Weibel instability, decreases with increasing relative density of the beam. We also found significant transverse magnetic component associated with the superluminal Langmuir turbulence, which has been mistakenly regarded as evidence of the F emission in earlier study. Further investigations are required to reveal their origin.
\end{abstract}


\section{Introduction} \label{sec:intro}

The plasma emission (PE) refers to electromagnetic radiation at frequencies close to the plasma oscillation frequency ($\wpe$) and its harmonics, corresponding to the fundamental (F) and harmonic (H) plasma emission, respectively. The framework of the standard theory of PE was suggested more than 6 decades ago \citep{1958SvA.....2..653G}, which involves a multi-step nonlinear process of wave-particle and wave-wave interactions due to the presence of energetic beam electrons in plasmas. The theory starts from the beam-driven kinetic bump-on-tail instability that leads to enhanced Langmuir waves, followed by the decay of these waves or their scattering over ion-related fluctuations generating the fundamental (F) emission and/or the backward-propagating Langmuir wave, and the nonlinear interaction of forward and backward-propagating Langmuir waves generating the harmonic (H) emission \citep[e.g.,][]{melrose_emission_1980,Melrose1987,1987JPlPh..38..169C,1994ApJ...422..870R, 2012JGRA..117.4106S,2014JGRA..119...69S, 2013JGRA..118.4748L, 2014SoPh..289..951L,2015A&A...584A..83T,2017PNAS..114.1502C,2019JGRA..124.1475H}.

The standard theory has been widely used to explain radio bursts in space and astrophysical plasmas, such as solar radio bursts in terms of type I-V \citep[see, e.g.,][for latest observational studies]{2014ApJ...787...59C,2016ApJ...830L...2V,2019ApJ...870...30V,2017SoPh..292...82L,2017SoPh..292..194L}, emissions from planetary electron foreshocks \citep[e.g.,][]{1984JGR....89.6631E,1984GeoRL..11..869M,1987JPlPh..38..169C,2005JGRA..110.7107K,2017EPSC...11..585P} and the outer heliosphere \citep{1984Natur.312...27K,1994JGR....9914729Z,1995AdSpR..16i.279G,1998GeoRL..25.4433G,1995GeoRL..22.3433C,1998ApJ...506..456C,2002GeoRL..29.1143C,2022ApJ...928..125T}. The theory involves evolution of turbulence and complicated multi-step nonlinear processes. A proper numerical verification of the theory requires kinetic simulations by the nature of the problem. Both ion-scale and electron-scale kinetic physics should be included. This excludes any fluid/hybrid models. Only fully-kinetic methods can be used, including the Vlasov method and the PIC method.

The Vlasov method directly solves the Vlasov-Maxwell equations to resolve the velocity distribution function of particles. The weak-turbulence (WT) theory/simulation represents a variety of this method tailored for the study of plasma emission \citep[e.g.,][see \citet{2019ApJ...871...74L} for the complete list of references]{2006PhPl...13b2302Y,2012PhPl...19j2303Y,2015ApJ...806..237Z}.

The PIC simulations numerically resolve the motion of each particle and the evolution of the electro-magnetic fields by solving self-consistently the equation of particle motion and the full-set of the Maxwell equations. In comparison to the WT method, the PIC method has the advantage of making no approximations to the basic laws of mechanics and electro-magnetism. We therefore choose the PIC method for the present study. See \citet{2019ApJ...871...74L} for a detailed comparison of the two methods when being applied to the problem of plasma emission. Earlier studies along this line of research include \citet{2001JGR...10618693K}, \citet{2007P&SS...55.2336K}, \citet{2009ApJ...694..618R,2009JKPS...54..313R}, \citet{2010JGRA..115.1204U}, \citet{2012ApJ...751..145G,2012SoPh..280..551G}, and latest studies include \citet{2015A&A...584A..83T}, \citet{2019JGRA..124.1475H}, \citet{2020ApJ...891L..25N}, and \citet{2022ApJ...924L..34C}. The PIC simulations demand massive computing and turns out to be challenging as
elaborated below.

\begin{enumerate}
\renewcommand{\labelenumi}{(\theenumi)}
  \item The number of macroparticles per cell per species (NPPCPS) should be large enough to lower the levels of numerical noise. Most earlier simulations have adopted values of NPPCPS less than a few hundred \citep[e.g.,][]{2001JGR...10618693K,2010JGRA..115.1204U} for economic reason. This results in relatively strong noise and low signal-noise ratio of the obtained F/H emissions since that are intrinsically weak. The situation becomes worse for weaker beam.
  \item The simulation domain should be large enough to have a good resolution of wave number ($k$). This is necessary for the F emission whose frequency upon excitation is close to the cutoff ($\sim \wpe$). Since according to the standard theory of PE the F emission is given by the scattering of the Langmuir (L) wave over the low-frequency ion-acoustic (IA) mode, thus its frequency is the sum or difference of the frequencies of the L and IA modes. The IA frequency is negligible compared to the Langmuir frequency ($\sim \wpe$), the frequency of the F emission is therefore very close to $\wpe$, which is the cutoff frequency of the fast electro-magnetic mode propagating in unmagnetized plasmas. It is therefore characterized by a long wavelength ($\lambda$) and small $k$. According to the simplified dispersion relation of electromagnetic wave ($\omega^2 = \wpe^2 + k^2 c^2$), with $kc \le  0.2~\wpe$ one gets $\lambda \ge 1500~\lde$ (the electron Debye length) for coronal plasmas at 1--2 MK (with a thermal speed of $\sim$ 0.02 c). The minimum dimension of the domain should be larger than a few wavelengths for proper simulation \citep[see, e.g.,][]{2019JGRA..124.1475H}.
  \item The temporal length of data used for the Fourier spectral analysis should be long enough to resolve modes with close frequencies. This is important since the F emission has almost the same frequency as the Langmuir mode. For instance, if the data length is taken to be $\sim$ 100 $\wpe^{-1}$, the two modes cannot be separated when their frequency difference is less than $0.1~\wpe$.
\end{enumerate}

The above elaborations point out the need of massive computation, while most earlier simulations did use insufficient number of macro-particles and limited domain. See the critical review given by \citet[][referred to as TT15 hereinafter]{2015A&A...584A..83T}. Note the domain is regarded as limited here if it is comparable to or only a few (say, $\sim $ 2--3) times larger than the the wavelength of the F emission. These limitations cause excessive numerical noise along dispersion curves and limited resolution of wave modes. The conclusions drawn are thus controversial regarding the generation of the F emission. For example, TT15 claimed that they presented ``the first self-consistent demonstration of F and H emission from a single-beam (unmagnetized) plasma system via fully kinetic PIC simulations'' --- in accordance with the standard theory of plasma emission. \citet{2019JGRA..124.1475H}, however, found that the F emission is hardly discernible according to their simulations with domain and NPPCPS being much larger than that of TT15. Thus, it remains open regarding whether the F emission does occur according to the standard theory (from the perspective of PIC verification), despite the positive statement of TT15 which is nevertheless inconclusive according to the following arguments.

\begin{enumerate}
\renewcommand{\labelenumi}{(\theenumi)}
  \item The domain dimension of TT15 is 600 $\lde$, much less than the expected wavelength of the F emission. According to TT15, the expected $k \lde$ of the F emission is about 0.002, corresponding to a wavelength of a few thousands of  $\lde$.
  \item The duration of data for the Fourier analysis is 100 $\wpe^{-1}$, the corresponding spectral resolution is less than $0.1~\wpe$, too poor to separate the F emission from nearby Langmuir waves with the obtained $\omega$--$k$ diagram.
  \item According to TT15, one evidence of the F emission is the growth of transverse component of magnetic field (see Figure 5 in TT15). Yet, such signal may have two contributions, according to our study presented here. One is given by the expected F emission, the other is by the Langmuir turbulence. The latter is found with $k \lde$ varying in a range of [0, 0.02], much broader than the expected $k \lde$ range of the F emission. Such signal is dominated by the Langmuir turbulence rather than by the F emission. This point cannot be inferred from earlier studies due to poor resolutions of both $\omega$ and $k$.
\end{enumerate}

Further, according to the dispersion relation of the fast electro-magnetic wave propagating in unmagnetized plasmas and the Faraday$'$s law, the relative strength of the wave E and B fields is determined by the ratio of $\omega^2 / k^2 c^2$. Thus, for the F emission with frequency ($\omega$) very close to $\wpe$, the E-field is much stronger than the B-field, so the F emission is dominated by the E-field. It is therefore critical to examine the E-field data to reveal its spectral characteristics and energy budget, to understand its generation mechanism.

In this study, we conduct the fully-kinetic PIC simulations in a domain much larger than that of TT15 to increase the spectral resolution, and with large value of NPPCPS to weaken the numerical noise. The main purpose is to verify the standard theory of plasma emission in unmagnetized plasmas energized by a single beam of energetic electrons. Signatures of the F emission and its generation process are highlighted to clarify existing contradiction in literature.

We first present the simulation method and parameter setup in the following section, together with the result for the thermal case, then present the cases with different number density ratio of beam and background electrons ($n_\mathrm{b}/n_\mathrm{0}$) in Sections 3 and 4. We adopt the same plasma and beam parameters for Case A as those in TT15 for a direct comparison. Cases with two values of $n_\mathrm{b}/n_\mathrm{0}$ (0.0057 and 0.05) have been modeled in TT15, while here an additional case with $n_\mathrm{b}/n_\mathrm{0} = 0.01$ is included to extend the analysis. The last section presents the conclusions and discussion.

\section{Parameter Setup and the Thermal Case}

Following our earlier studies of coherent emission in magnetized plasmas \citep[e.g.,][]{2020ApJ...891L..25N,2021ApJ...909L...5L,2021ApJ...909....3Y,2021ApJ...920L..40N,2022ApJ...924L..34C}, we continue to use the Vector-PIC \citep[VPIC:][]{5222734,2008PhPl...15e5703B,Bowers_2009} code released by the Los Alamos National Laboratory (LANL), which are 2D3V, i.e., two-dimensional in space and three-dimensional for particle velocity and electromagnetic fields. We present three cases (A, B, and C) with different $n_\mathrm{b}/n_\mathrm{0}$ (= 0.0057, 0.01, and 0.05) within simulation domain as large as $\sim$ 6000 $\lde$ $\times$ 6000 $\lde$. The cell size is set to be 2.929 $\lde$ and the number of cells to be 2048 along both $\hat{e}_x$ and $\hat{e}_z$ direction, where $\hat{e}_x$ ($\hat{e}_z$) represents the unit vector along the $x$ ($z$) direction. The beam is along $\hat{e}_z$, and $\vec k$ lies in the $xOz$ plane. The background electron-proton plasma is assumed to be unmagnetized and thermal, to compare directly with earlier studies. The time step is set to be 0.03 $\wpe^{-1}$. The NPPCPS are 4000 for background electrons, and 1000 for both background protons and beam electrons. The total number of macroparticles is 2.4 $\times~10^{10}$. In Table A1 of the Appendix, we have presented some setup parameters of several PIC studies on plasma emission for comparison. According to the table, the present study has almost the same size of simulation domain as that used by its sister paper \citep{2022ApJ...924L..34C} and the largest NPPCPS and total number of macroparticles among the listed studies.

The details of the simulation setup with the domain dimension, cell size, NPPCPS, resolvable ranges of wavenumber ($k$) and frequency ($\omega$) have been listed in Table 1. Parameters used by TT15 are also included. To illustrate the effect of the domain size, we include another case (Case A$^\prime$) within a smaller domain (1200 $\times$ 1200 $\lde^2$). Other parameters of Case A$^\prime$ are the same as Case A.

The beam electrons are represented with the following drifting Maxwellian distribution function
\begin{equation}
f_\mathrm{b} = A_\mathrm{b} \exp(-\frac{u^2_\perp}{2u^2_{0}} - \frac{(u_\parallel - u_\mathrm{d})^2}{2u^2_{0}})
\end{equation}
where $u_\parallel$ and $u_\perp$ are the parallel and perpendicular components of the momentum per mass, $u_d$ is the average drift momentum per mass of the beam electrons, $u_0$ is the thermal velocity of energetic electrons, and $A_\mathrm{b}$ is the normalization factor. The parameters for the background plasmas and the beam are taken from TT15, which are 2 MK ($\sim 0.02c$) for $T_\mathrm{e}$, $T_\mathrm{p} = 0.7~T_\mathrm{e}$, and 0.3c for the average drift speed of the beam and the beam temperature is set to be $T_\mathrm{e}$.

To tell the significance of plasma emission, it is essential to compare their spectral intensity with that given by the thermal case (Case T), in which wave-like noises exist along dispersion curves of intrinsic wave modes even without any energetic electrons. Other setup parameters of Case T are the same as those of Cases A--C. The modes attributed to energetic electrons must be considerably stronger than the corresponding numerical noise.

We start from analyzing Case T. See Figure A1 in the Appendix for the obtained $\omega$--$k$ diagram for Case T. To show the effect of NPPCPS we aslo present in Figure A1 another thermal case (T$^\prime$) with NPPCPS = 500 (for electrons) and 125 (for protons). The diagrams are along two propagating directions with $\theta = 30^\circ$ and $\theta = 80^\circ$. The color bar of the lower panels is 20 dB less than that of the upper two panels to show out weaker signals. Range of the normalized wavenumber ($k \lde$) for the uppermost panel is [$-$0.3, 0.3], while it is [$-$0.03, 0.03] for the lower three panels. Two conclusions can be drawn: (1) the numerical noise level is inverse-proportional to the NPPCPS, i.e., it is $\sim 9$ dB stronger in Case T$^\prime$ than in Case T; (2) there exist two high-frequency modes in unmagnetized plasmas (as expected), the electrostatic Langumuir mode and the fast electromagnetic mode. The Langmuir mode (the upper two panels) extends to frequencies above $\wpe$ due to the thermal effect, which is well-described by the following relation,
\begin{equation}
\omega^2 = \wpe^2 (1 + 3 k^2 \lde^2)
\end{equation}
The Langmuir wave exists continuously from superluminal to subluminal regimes, being perfectly electrostatic. The transverse mode is also dominated in energy by the $E_y$ and $E_z$ components that are stronger by about 10--20 times than the magnetic counterparts ($B_x$ and $B_y$). For Case T, the maximum intensity of $E_y$ is about -140 dB and $B_y$ about -150 dB, as read from the figure. These values represent the levels of modes relevant to numerical noises.

\section{Simulation Results for Case A and Comparison With TT15}

Figure 1 and the accompanying movie present the temporal evolution of the velocity distribution function (VDF) of electrons for Case A with $n_\mathrm{b}/n_\mathrm{0} = 0.0057$. Due to the rapid growth of the bump-on-tail instability, the beam electrons decelerate rapidly and diffuse towards the regime with lower $v_\parallel$. This occurs during the first few hundred $\wpe^{-1}$. The beam electrons also diffuse towards the regime with larger $v_\perp$, indicating efficient perpendicular heating. Such heating is induced by two processes, the perpendicular diffusion due to the bump-on-tail instability \citep[see][]{2020PhPl...27b0702H,2021SoPh..296...42M,2022ApJ...924L..34C} and the electromagnetic Weibel instability that was discovered by \citet{1959PhRvL...2...83W} for anisotropic plasmas with bi-Maxwellian distribution and by \citet{1959PhFl....2..337F} for plasmas with counter-streaming electrons. In our case, the instability is driven by the single electron beam, leading to the growth of the electromagnetic beam mode with significantly-enhanced $B_y$ component and perpendicular heating of the beam electrons \citep[see also][TT15]{2009ApJ...690..189K}. The VDF reaches the asymptotic state after $\sim$ 1000 $\wpe^{-1}$.

In Figure 2a, we show energy curves of the six field components and the negative change of the total kinetic energy of all electrons ($-\Delta E_k$) for Cases T and A. The beam-plasma interaction causes rapid rise of $E_z, E_x$, and $B_y$ within the first 80 $\wpe^{-1}$, while the other field components remain at the background levels. The maximum intensity of $E_z$ reaches about 13$\%$ of $-\Delta E_k$, while those of $E_x$ and $B_y$ reach about 2$\%$ and 0.01$\%$ of $-\Delta E_k$, respectively. As seen from the following analysis, the $E_x$ and $E_z$ components mainly correspond to the Langmuir mode while the $B_y$ component is energized by the low-frequency electromagnetic Weibel instability. The $E_x$ and $E_z$ intensities maintain an almost constant level before declining gradually after $\sim$ 1000 $\wpe^{-1}$, representative of the nonlinear evolution of turbulent plasmas in quasi-equilibrium.

In Figure 2b we have plotted the energy of field components associated with various wave modes. The temporal variation of mode energy is given by the inverse Fourier transform of the total spectral energy within the selected spectral regimes (see Figure A2). Among various wave modes, the strongest one is the BL mode given by the coupling of the beam mode and the intrinsic Langmuir mode. Its $k$ range is [0.08, 0.2] $\lambda_\mathrm{De}^{-1}$, and frequency range is [0.9, 1.3] $\wpe$. It propagates from $\theta = 0^\circ$ (i.e., along the beam) to $\theta = 60^\circ$. The energy curve of $E_z$ (Figure 2a) agrees with the Langmuir mode energy (Figure 2b).

We performed the exponential fittings to the energy profiles of the $E_z$ field and the beam-Langmuir (BL) mode (see the dashed lines overplotted in Figure 2). The obtained linear growth rate is 0.076 $\wpe^{-1}$ for $E_z$, and 0.053 $\wpe$ for the BL mode energy. The theoretical growth rate $\gamma$ and the corresponding frequency $\omega_\mathrm{r}$ versus $k \lde$, according to Equation (1) of the Appendix, has been plotted in the upper panels of Figure A3. The parameters used for the plots are identical to those of Case A. In Figure A3a, the maximum growth rate of the bump-on-tail instability ($\gamma_\mathrm{M}$) is 0.094 $\wpe$ at ($\omega_\mathrm{r}$, $k\lde$) = (0.96 $\wpe$, 0.069), and the range of significant growth (within which the growth rate is larger than, say,  $\omega_\mathrm{rM}/2$) extends from $(\omega_\mathrm{r}$, $k\lde$) = (0.75 $\wpe$, 0.049) to (1.06 $\wpe$, 0.084). On the other hand, according to the PIC simulation (see Figure A3b for the dispersion diagram of Case A within the period of [0, 500 $\omega_\mathrm{pe}^{-1}$]), the BL mode presents the maximum growth around $(\omega_\mathrm{r}$, $k\lde$) = (0.98 $\omega_\mathrm{pe}$, 0.08), and the range of significant growth extends from $(\omega_\mathrm{r}$, $k\lde$) = (0.98 $\wpe$, 0.075) to (1.01 $\wpe$, 0.12). We concluded that the PIC simulation agrees reasonably well with the theoretical prediction.

Figure 3 presents the wave map in the $\vec k$ space. Figure 4 presents the corresponding $\omega$--$k$ diagrams along $\theta = 30^\circ$ and/or $80^\circ$ for the modes with frequencies close to $\wpe$, and Figure 5 presents the similar diagrams for the modes at the low frequency ($\ll  \wpe$) and the harmonic frequency ($\sim 2 \wpe$). These figures (and accompanying movies) shall be combined to reveal the characteristics of each mode. According to Figures 3 and 4, there exist two other Langmuir components on the left (or backward) side of BL, referred to as the generalized Langmuir (GL) mode in total. One is forward-propagating with a narrow range of k ([-0.04, 0.04] $\lambda_\mathrm{De}^{-1}$), the other is backward-propagating with a larger range of k ([-0.2, -0.04] $\lambda_\mathrm{De}^{-1}$).  The forward one is mostly superluminal. According to the movie accompanying Figure 3 and the energy curves of Figure 2, the backward one appears shortly after the onset of BL, while the forward one appears after 100 $\wpe^{-1}$. The two components have very similar energy profiles, both reaching the maximum around 1000 $\wpe^{-1}$, indicating the same physical origin.

From panels (b)--(d) of Figure 4, we observe significant enhancements along the corresponding dispersion curve of the F emission, i.e., the fast transverse electromagnetic mode around $\wpe$, from the $\omega$--$k$ diagrams of $E_z$ and $B_y$. According to Figure 2b, its energy (summed over all propagating directions) reaches about $10^{-5}~E_{k0}$, a fraction of the total energy of the H emission. As expected, the F emission is dominated in energy by its $E_z$ component, which is stronger than $B_y$ by about 2--3 orders in magnitude. It reaches the maximum intensity around 1000 $\wpe^{-1}$, with a variation trend similar to the H emission.

The harmonic (H) emission emerges as the circular feature after $\sim$ 500 $\wpe^{-1}$ in the maps of $E_z$, $E_x$, and $B_y$ (see Figures 3c and 5c, and accompanying movies), and saturates at the level of $10^{-5}~E_{k0}$ after 1000 $\wpe^{-1}$. Within the H circle, there appears enhanced quadrupolar $B_y$ feature which is mainly generated by the beam-driven electromagnetic Weibel instability at low frequency (see Figure 5b, see also TT15). The $B_y$ signal rises rapidly during the initial stage of the plasma-beam interaction, reaching the maximum intensity around 100 $\wpe^{-1}$.

Figure 5a presents the ion-acoustic (IA) mode that is very weak due to the strong Landau damping in plasmas with $T_\mathrm{p}/T_\mathrm{e}= 0.7$. See the lower two panels of Figure A3 in the Appendix for the frequency and growth rate of the IA mode within the background Maxwellian plasmas, evaluated according to Equation (2) in the Appendix.

Another interesting feature is the weak yet significant $B_y$ enhancement along the Langmuir dispersion curve (see Figure 4c). It occupies a much larger range of $k$ ([-0.04, 0.04] $\lambda_\mathrm{De}^{-1}$), with the spectral intensity weaker than the F emission. Yet, the total energy of the Langmuir $B_y$ component is much larger than that of the F emission. It reaches the maximum level ($\sim 10^{-6}~E_{k0}$) also around $1000~\wpe^{-1}$ (see Figure 2c), $\sim 10$ times stronger than the simultaneous F emission. Similar $B_y$ enhancement is also observed from Figure 5 of TT15. Yet, due to the poor spectral resolution there it is not possible to separate the $B_y$ component of the F emission from the prevailing Langmuir enhancement. Such $B_y$ enhancement is not associated with the Langmuir noise of thermal plasmas according to Figure A1.

To support the above argument, we conducted another simulation (Case A$^\prime$) within a much-smaller domain (1200 $\times$ 1200 $\lde^2$). See Table 1 for the simulation setup. The plasma-beam parameters are not changed. The obtained wave map in the $\vec k$ space is shown in Figure A4 of the Appendix for large (panel a) and small (panels b and c) range of $k$. The spectra of the three magnetic field components are presented in panel c. The obtained $\omega$--$k$ diagrams are presented in Figures A5 of the Appendix, along two propagating angles with $\theta = 30^\circ$ and $\theta = 80^\circ$. The accompanying movie presents the complete dispersion diagrams from $\theta = 0^\circ$ to $\theta = 90^\circ$. For direct comparison, the data duration used in the Fourier analysis  is taken to be 100 $\wpe^{-1}$, the same as that used by TT15. We see that the circular H emission cannot be identified clearly, and there exist no regular wave enhancements along the dispersion curve of the F emission, no way to tell the significance of its $B_y$ field due to the presence of the overwhelming Langmuir $B_y$ component. Note that TT15 has mistakenly regarded the total $B_y$ enhancement to be the evidence of the F emission.

\section{Simulation Results for Cases B and C and Comparison with Case A}

Figures 6 and 7 present the wave distribution map in the $\vec k$ space and the $\omega$--$k$ diagram for Cases B ($n_\mathrm{b}/n_\mathrm{0} = 0.01$) and C ($n_\mathrm{b}/n_\mathrm{0} = 0.05$). The energy curves for Case B have been plotted in Figures 2c--2d. They should be combined with the results for Case A to understand the effect of varying $n_\mathrm{b}/n_\mathrm{0}$. Note that the illustrated spectral energy is `relative' since it has been normalized by the total initial beam energy ($E_{k0}$) of each individual case. $E_{k0}$ increases with increasing $n_\mathrm{b}/n_\mathrm{0}$, being $4.12~m_\mathrm{e} c^2$ for A, $7.25~m_\mathrm{e} c^2$ for B, and $37.7~m_\mathrm{e} c^2$ for C.

The most obvious result is that all modes discussed here, except the low-frequency electromagnetic Weibel mode, decrease in relative spectral energy with increasing $n_\mathrm{b}/n_\mathrm{0}$. All modes look very similar when $n_\mathrm{b}/n_\mathrm{0}$ increases from 0.0057 to 0.01, yet, when $n_\mathrm{b}/n_\mathrm{0}$ further increases to 0.05, only the forward BL mode together with a small part of the forward propagating GL wave remain while the backward one and the F/H emissions vanish. The Weibel instability responds to increasing $n_\mathrm{b}/n_\mathrm{0}$ differently, yielding stronger $B_y$ in terms of relative spectral energy. Further discussion on the Weibel instability is beyond the scope of the study.

As observed from Figure 7, the central frequency of BL in Case C is about 0.86 $\wpe$, being considerably lower than in Cases A and B. The difference from $\wpe$ is too large to allow the standard plasma emission process to occur. This agrees with TT15 that is based on simulations within a much smaller domain.

\section{Conclusions and Discussion}
Our main purpose is to verify the standard model of plasma emission which describes a multi-stage nonlinear process in the aftermath of the kinetic bump-on-tail instability of a single-beam plasma system. The primary outcome of the instability is the enhanced turbulence of electrostatic beam-Langmuir (BL) mode, which further interacts with other secondary fluctuations to yield the F and H radiations. Previous verification studies using PIC simulations, limited by available computational resources, have drawn contradictory statements regarding the significance of the F emission. In this study we employed a simulation domain and number of macroparticles that are among the largest ones of similar studies, to lower noise levels and achieve higher spectral resolution.

We found that for the ratio of number density ($n_\mathrm{b}/n_\mathrm{0}$) being less than 0.01, significant F emission can be generated. The F emission reaches an energy level of 10$^{-5}$ of the total initial beam energy $E_{k0}$ that is a fraction of the total energy of the H emission. The spectral energies of all wave modes relative to $E_{k0}$, except the beam-driven electromagnetic Weibel instability, decrease with increasing $n_\mathrm{b}/n_\mathrm{0}$. In agreement with TT15, for $n_\mathrm{b}/n_\mathrm{0} \approx 0.05$ the frequency of the BL mode is too low to meet the matching conditions of three-wave interaction, thus the backward-propagating Langmuir wave and the F and H emissions cannot be generated.

According to the simulation result, the IA mode is indeed extremely weak in plasmas with $T_\mathrm{p}/T_\mathrm{e}= 0.7$ due to the strong Landau damping effect (see Figure A3). Thus, the most promising process is the `d' or the decay process (BL $\rightarrow$ F + IA), rather than the `u' process (BL + IA $\rightarrow$ F). Due to the thermal effect, the maximum frequency of the BL mode is about 1.04 $\wpe$, exceeding the cutoff of the F emission by 0.04 $\wpe$. This also supports the occurrence of the `d' process, indicating that the thermal correction to the dispersion relation of Langmuir wave is critical to the emission process.

In the case with $n_\mathrm{b}/n_\mathrm{0} \le 0.01$, we found significant enhancement of the transverse component $B_y$ at frequency around $\wpe$ and in a $k$ range of [-0.02, 0.02] $\lambda_\mathrm{De}^{-1}$, with contributions from both the F emission and the beam-Langmuir turbulence. It remains unresolved regarding the generation mechanism of such Langmuir $B_y$ component (that is about three orders in magnitude weaker than the corresponding $E_z$ component). Note that in the simulation with only thermal plasmas the Langmuir noise manifests no signatures of $B_y$ (see Figure A1). Future study should explore how such $B_y$ component develops in such plasmas of turbulent equilibrium.

Using the same PIC code, \citet{2022ApJ...924L..34C} examined the plasma emission process in weakly magnetized plasmas ($\wpe/\Omega_\mathrm{ce} = 10$) interacting with a single beam of energetic electrons. They presented cases with arbitrarily different mass ratio ($m_\mathrm{p}/m_\mathrm{e}$) and found that with increasing $m_\mathrm{p}/m_\mathrm{e}$ the intensity of BL increases correspondingly, together with enhancements of Z mode and F/H emissions. This indicates that the latter modes (Z and F/H) are secondary products of the primary BL mode. This agrees with the study presented here.

Comparing the case for weakly-magnetized background plasmas \citep{2022ApJ...924L..34C} and Cases A and B presented here for unmagnetized plasmas, we find that the two sets of solution are quite similar. In particular, they are similar in the overall temporal evolution of the EVDF, the growth and characteristics of the beam-Langmuir wave, and backward and forward-propagating secondary generalized-Langmuir waves. The main difference between the two cases lies in the generation of the superluminal Z mode (with cutoff frequency below $\wpe$) and the whistler mode in \citet{2022ApJ...924L..34C} while in this study there present the generalized Langmuir wave and the electromagnetic mode induced by the beam-Weibel instability. These major similarities support the standard theory of plasma emission works for the two cases, that is, the electromagnetic decay to generate the F emission, and nonlinear coalescence of the BL wave with the backward-propagating Langmuir wave to generate the H emission.

The above argument does not support the plasma emission mechanism suggested by \citet{2020ApJ...891L..25N,2021PhPl...28d0701N} that the F emission is generated by almost-counterpropagating Z and W modes and the H emission by almost-counterpropagating upper-hybrid (UH) mode. Their conclusion is based on PIC simulations of plasmas interacting with trapped energetic electrons with the loss-cone type VDFs. There the primary mode is the UH mode which is the Langmuir wave with a large propagating angle relative to the background magnetic field. The UH mode and the secondary Z and W modes are excited directly through the electron cyclotron maser instability (ECMI). This suggests that the mechanism of plasma emission may be different for different VDFs of energetic electrons: for beam electrons the standard plasma emission mechanism is at work, while for trapped electrons the plasma emission induced by ECMI may be important.

\begin{acknowledgments}
This study is supported by NNSFC grants (11790303 (11790300), 11973031, and 11873036). The authors acknowledge the Beijing Super Cloud Computing Center (BSCC, URL: \url{http://www.blsc.cn/}) for computational resources, and LANL for the open-source VPIC code.
\end{acknowledgments}

\appendix
Here we present the $\omega$--$k$ diagram for the two thermal cases (T and T$^\prime$, with only Maxwellian plasmas) in Figure A1, with different NPPCPS to show its effect on the numerical noise level. In Case T the NPPCPS = 4000 (for electrons) and 1000 (for protons); in Case T$^\prime$ the NPPCPS = 500 (for electrons) and 125 (for protons). The diagrams are along two propagating directions with $\theta = 30^\circ$ and $\theta = 80^\circ$.

In Figure A2, we present the spectral regimes selected to evaluate the mode energy as plotted in Figure 2.

In Figure A3a, we plot the growth rate ($\gamma$) and frequency ($\omega_\mathrm{r}$) of the beam-Langmuir mode excited via the kinetic bump-on-tail instability, according to the following equation
\begin{equation}
1+\frac{2 \omega_\mathrm{p e}^{2}}{k_{\|}^{2} v_\mathrm{T e}^{2}}\left(1+\xi_\mathrm{e} Z\left(\xi_\mathrm{e}\right)\right)+\frac{2 \omega_\mathrm{p b}^{2}}{k_{\|}^{2} v_\mathrm{T b}^{2}}\left(1+\xi_\mathrm{b} Z\left(\xi_\mathrm{b}\right)\right)=0
\end{equation}
where $v_\mathrm{Te}$ and $v_\mathrm{Tb}$ represent the thermal velocity of the background and the beam electrons, $Z$ is the plasma dispersion function, and $\xi_\mathrm{e} = \frac{\omega}{k v_\mathrm{Te}}$,  $\xi_\mathrm{b} = \frac{\omega - k v_\mathrm{d}}{k v_\mathrm{Tb}}$, and $v_\mathrm{d}$ is the average drift velocity of the beam electrons.

In Figure A3b, we present the dispersion diagram of Case A within the period of [0, 500 $\omega_\mathrm{pe}^{-1}$]) to compare with the theoretical predictions of the ranges of $\omega$ and $k$ of significant beam-Langmuir wave growth. A shorter period (than that used in Figure 4) is employed here to avoid the influence of the nonlinear stage of wave evolution after $t=500$ $\omega_\mathrm{pe}^{-1}$.

In Figures A3c and A3d, we plot the growth rate ($\gamma$) and frequency ($\omega_\mathrm{r}$) of the IA mode, versus the normalized wave number and the proton-electron temperature ratio, according to
\begin{equation}
\omega_\mathrm{r}^2 = k^2 v_\mathrm{s}^2, \frac{\gamma}{\omega_\mathrm{r}}=-\sqrt{\frac{\pi}{8}}\frac{v_\mathrm{s}^3}{c_\mathrm{s}^3}\left[\sqrt{\frac{m_\mathrm{e}}{m_\mathrm{p}}}e^{-(\frac{v_\mathrm{s}}{v_{T_\mathrm{e}}})^2}+(\frac{T_\mathrm{e}}{T_\mathrm{p}})^{\frac{3}{2}}e^{-\frac{3}{2}}e^{-\frac{T_\mathrm{e}/T_\mathrm{p}}{2(1+k^2\lde^2)}}\right]
, v_\mathrm{s}^2 = \frac{3}{2} v_{T_\mathrm{p}}^2 + \frac{c_\mathrm{s}^2}{ 1 + k^2 \lde^2} , c_\mathrm{s}^2 = \frac {k_\mathrm{B} T_\mathrm{e}}{m_\mathrm{p}}
\end{equation}
where $v_\mathrm{s}$ is the phase speed of the IA mode. According to Figures A3c and A3d, for $T_\mathrm{p}/T_\mathrm{e} = 0.7$ and $k \lde = 0.2$ (see Figure 5 for these typical parameters), we get $\omega_\mathrm{r} = 0.008~\wpe$ and $\gamma = 0.72~\omega_\mathrm{r} \approx 0.006~ \wpe$, then the e-folding time of the decay of IA is $\tau = \frac{1}{\gamma} \approx 170~\wpe^{-1} \approx 0.2~T$, where the period $T = \frac {2\pi}{\omega_\mathrm{r}} \approx 780~\wpe^{-1}$. This means that the IA mode revealed here is heavily damped over just one period.

In Figures A4 and A5, we show the wave distribution map in the $\vec k$ space and the $\omega$--$k$ diagram for Case A$^\prime$ which has a much smaller simulation domain (1200 $\times$ 1200 $\lde^2$) than Case A. The other setup parameters of the two cases are taken from Case A.

\bibliography{Zhang_ZL_2022ApJ}{}
\bibliographystyle{aasjournal}

\begin{deluxetable*}{cccccccc}
\tablenum{1}
\tablecaption{Setup parameters of simulation, including the domain dimension, the cell size, the time step, the duration, resolutions of $k$ and $\omega$, and NPPCPS. Parameters used by TT15 are also listed.}
\tablewidth{0pt}
\tablehead{
\colhead{Case \#}  & \colhead{Domain} & \colhead{Cell size / Time step} & \colhead{Duration} & \colhead{$[k_\mathrm{m},k_\mathrm{M}]^{a}$} & \colhead{$[\omega_\mathrm{m},\omega_\mathrm{M}]^{b}$} & \colhead{$[\omega_\mathrm{m},\omega_\mathrm{M}]^{c}$}   & \colhead{NPPCPS$^{d}$}\\
\colhead{}  & \colhead{[*$\lambda_\mathrm{De}$, *$\lambda_\mathrm{De}$]} & \colhead{$\lambda_\mathrm{De}$ / $\omega_\mathrm{pe}^{-1}$} & \colhead{($\omega_\mathrm{pe}^{-1}$)} & \colhead{($\lambda_\mathrm{De}^{-1}$)} & \colhead{($\omega_\mathrm{pe}$)} & \colhead{($\omega_\mathrm{pe}$)}   & \colhead{}
}
\startdata
Case A$^{\prime}$   & [1200, 1200] & $1.171/0.0116$ & 2000 & [0.005, 2.6] & [0.06, 3.14] & [0.006, 3.14] & 4000, 1000, 1000\\
Cases A--C, T & [6000, 6000] & $2.929/0.03$ & 2000 & [0.001, 1.07] & [0.06, 3.14] & [0.006, 3.14] & 4000, 1000, 1000\\
TT15                & [600, 600] & 1/NA & 1000 & [0.01, 3.1] & [0.06, NA] & NA & 1000, 1000, 1000
\enddata
\tablecomments{`a' represents  the resolvable range of $k$ given by $[2\pi/L_x, \pi/\delta_x]$, where $L_x$ represents the length of the domain, and $\delta_x$ represents the cell size.
`b' represents the resolvable range of $\omega$ with Fourier analysis over data duration of 1000 $\omega_\mathrm{pe}^{-1}$,
     given by $[2\pi/T, \pi/\delta_t]$, where $T$ represents the duration of the data, and $\delta_t$ represents
     the time step of the output data. The timestep to output the field data is 1.0 $\omega_\mathrm{pe}^{-1}$.
`c' represents the resolvable range of frequency ($\omega$) with Fourier analysis over data duration of 1000 $\omega_\mathrm{pe}^{-1}$.
`d' (NPPCPS) represents number of particles per cell per species (background electrons and protons, beam electrons).
}
\end{deluxetable*}

\begin{figure*}[ht]
 \centering
 \includegraphics[width=0.7\linewidth]{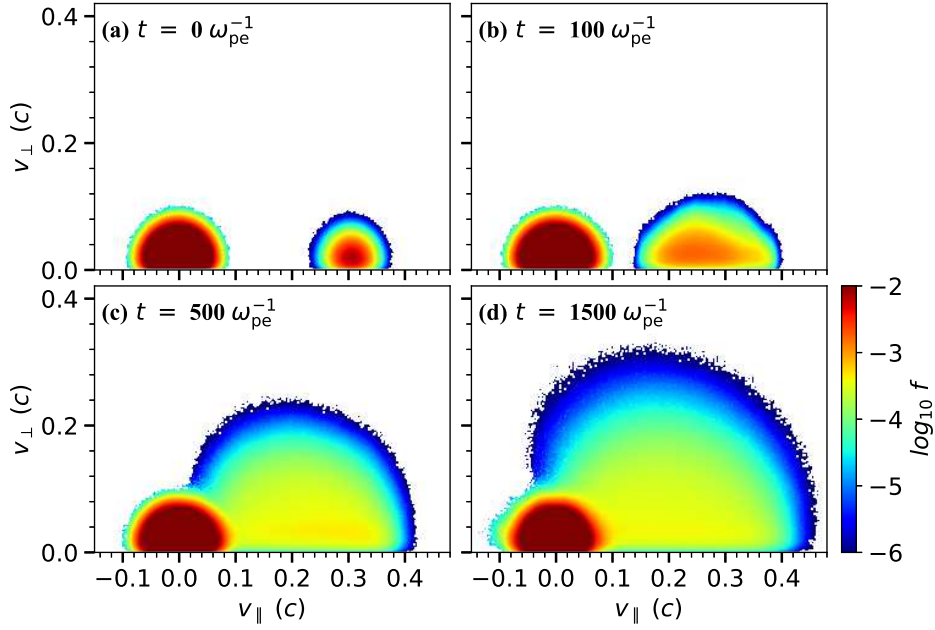}
  \caption{Temporal evolution of the VDFs at four representative moments. The video begins at $t=0~\wpe^{-1}$ and advances $100~\wpe^{-1}$ per frame till $t=2000~\wpe^{-1}$. The real-time duration of the video is 5 s. (An animation of this figure is available).}\label{fig1}
\end{figure*}

\begin{figure*}[ht]
 \centering
 \includegraphics[width=0.7\linewidth]{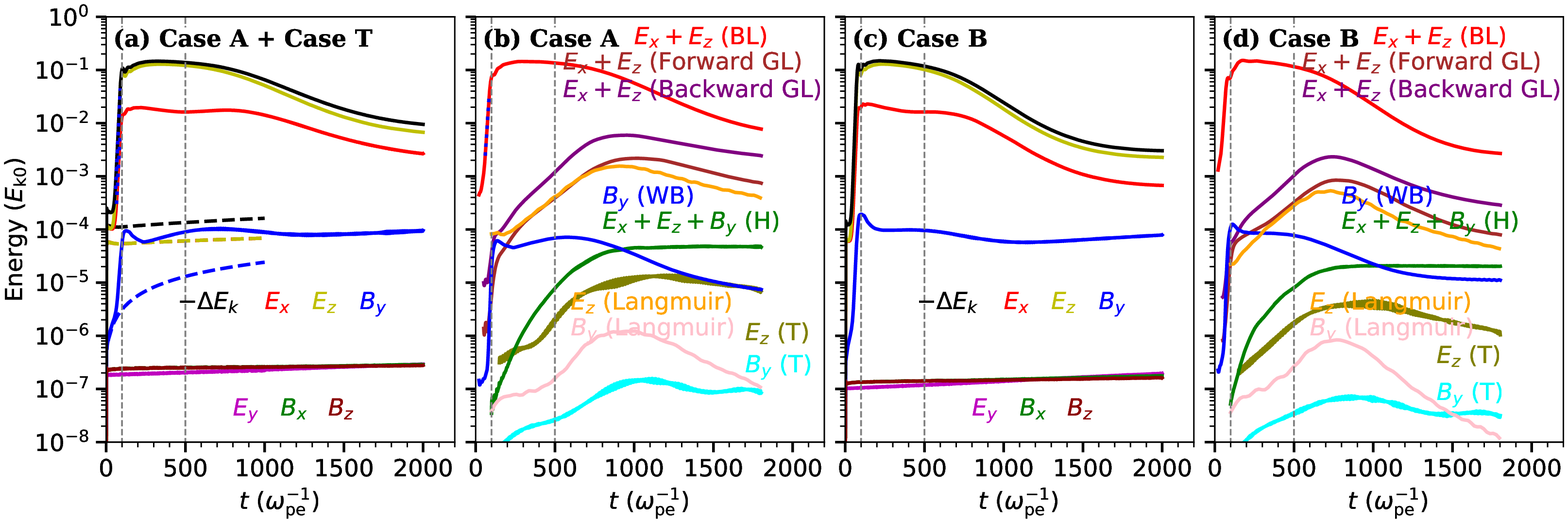}
  \caption{(a) Energy variation of electromagnetic fields and the negative change of the total electron energy ($-\Delta E_k$) for Case A; (b) energy variation of field components of various wave modes for Case A (BL for the beam-Langmuir mode, GL for the backward and forward- propagating generalized Langmuir mode, WB for the mode excited by the Weibel instability, F for the fundamental and H for the harmonic emissions). The spectral regimes adopted to evaluate these energy profiles are presented in Figure A2 of the Appendix. The overplotted dashed lines in panels (a) and (b) are given by the exponential fittings to the energy profiles of $E_z$ and the forward generalized Langmuir mode.}\label{fig2}
\end{figure*}

\begin{figure*}[ht]
 \centering
 \includegraphics[width=0.7\linewidth]{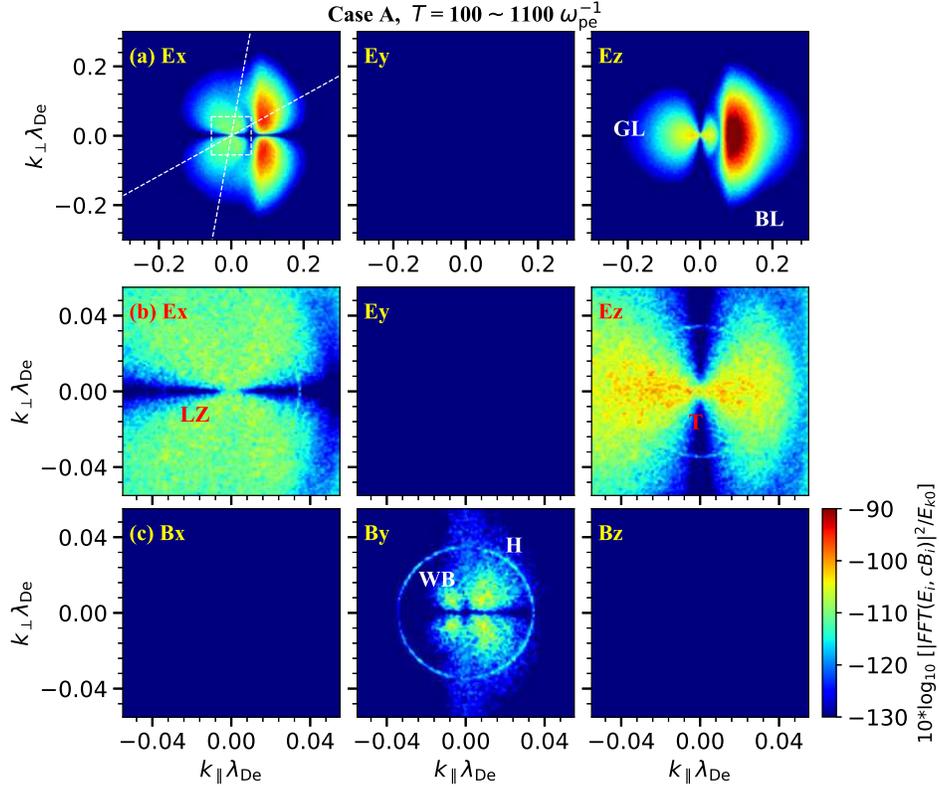}
  \caption{Wave intensity maps of the six field components in the wave-vector $\vec k$ space for Case A. The dashed lines in panel (a) represent $\theta = 30^\circ$ and $80^\circ$. Middle and lower panels are the zoom-in view of the white square drawn in panel (a). The energy is normalized by the total energy of beam electrons $E_{k0}$ of Case A. The video begins at $t=0~\wpe^{-1}$ and advances $20~\wpe^{-1}$ per frame till $t=200~\wpe^{-1}$ and then advances $100~\wpe^{-1}$ till $t=2000~\wpe^{-1}$. The real-time duration of the video is 5 s. (An animation of this figure is available).}\label{fig3}
\end{figure*}

\begin{figure*}[ht]
 \centering
 \includegraphics[width=0.6\linewidth]{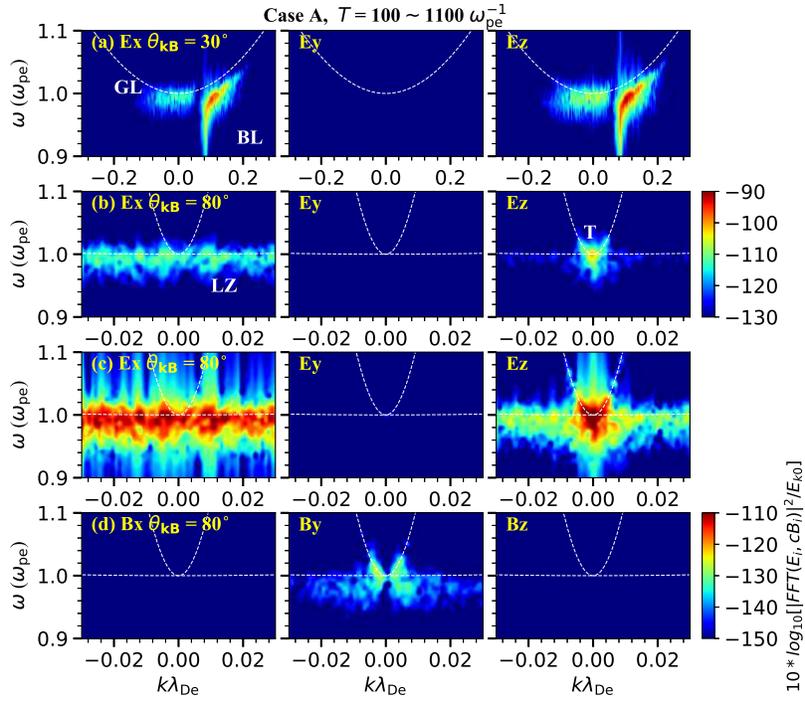}
  \caption{Dispersion diagrams of ($E_x, E_y, E_z$) and ($B_x, B_y, B_z$) for various wave modes in the frequency range of 0.9--1.1 $\wpe$ in large (panel a) and small (panels b--d) ranges of $k$. Analytical dispersion curves of corresponding wave modes are over-plotted. The video begins at $\theta = 0^\circ$ and advances $5^\circ$ at a time up to $\theta = 90^\circ$. The real-time duration of the video is 5 s. (An animation of this figure is available).}\label{fig4}
\end{figure*}

\begin{figure*}[ht]
 \centering
 \includegraphics[width=0.6\linewidth]{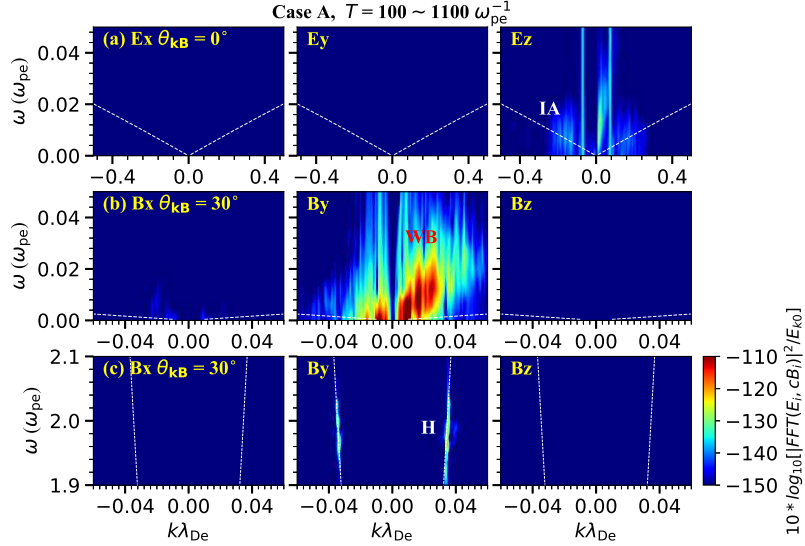}
  \caption{Dispersion diagrams of ($E_x, E_y, E_z$) and ($B_x, B_y, B_z$) for various wave modes in the frequency range of [0., 0.05] $\wpe^{-1}$ (panels a--b) and [1.9, 2.1] $\wpe^{-1}$ (panel c), for Case A. Analytical dispersion curves of corresponding wave modes (IA for ion acoustic mode) are over-plotted. The video begins at $\theta = 0^\circ$ and advances $5^\circ$ at a time up to $\theta = 90^\circ$. The real-time duration of the video is 5 s. (An animation of this figure is available).}\label{fig5}
\end{figure*}

\begin{figure*}[ht]
 \centering
 \includegraphics[width=0.99\linewidth]{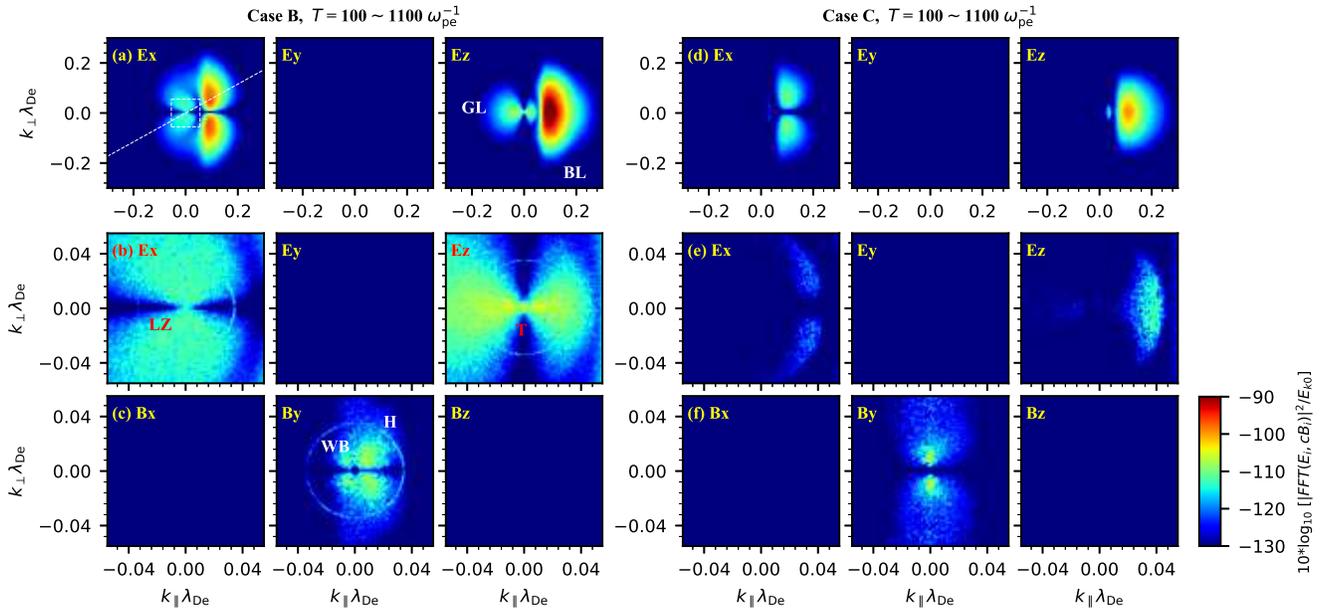}
  \caption{The same as Figure 3 yet for Case B (left panels) and Case C (right panels). The energy is normalized by the total energy of beam electrons $E_{k0}$ of the corresponding case. The video begins at $t=0~\wpe^{-1}$ and advances $20~\wpe^{-1}$ per frame till $t=200~ \wpe^{-1}$ and then advances $100~\wpe^{-1}$ till $t=2000~\wpe^{-1}$. The real-time duration of the video is 5 s. (An animation of this figure is available).}
\end{figure*}

\begin{figure*}[ht]
 \centering
 \includegraphics[width=0.99\linewidth]{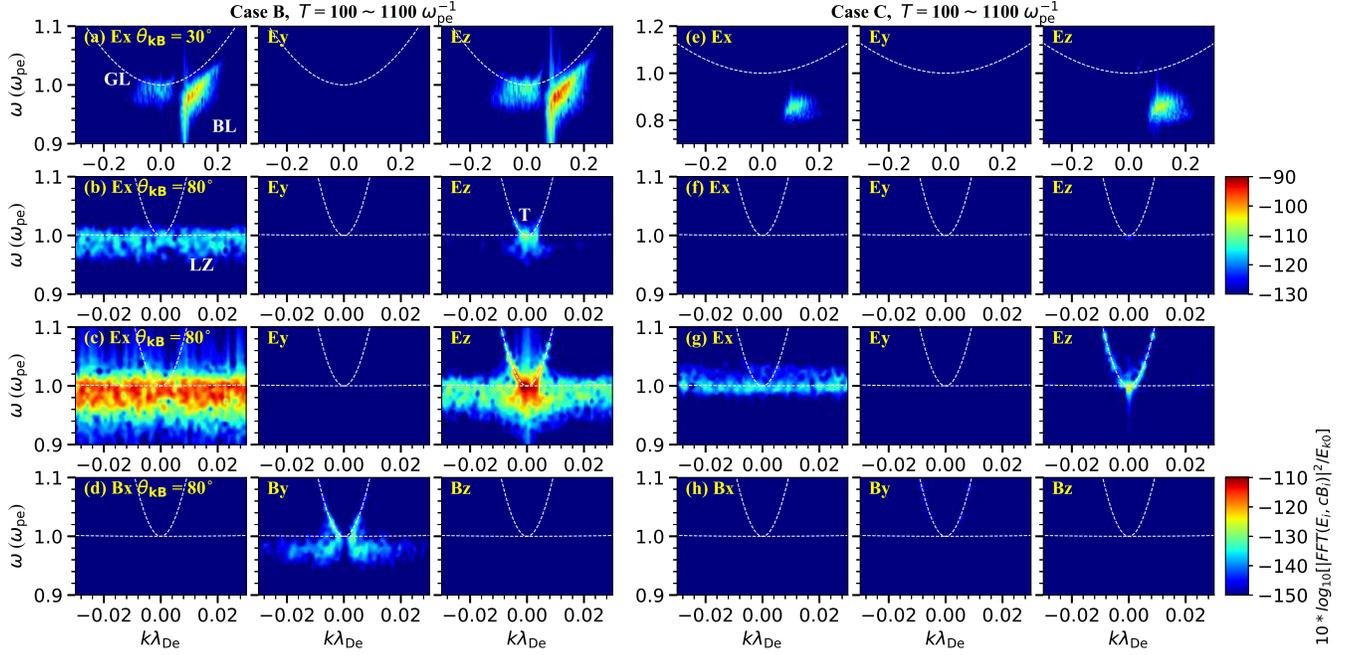}
  \caption{The same as Figure 4 yet for Case B (left panels) and Case C (right panels). The video begins at $\theta = 0^\circ$ and advances $5^\circ$ at a time up to $\theta = 90^\circ$. The real-time duration of the video is 5 s. (An animation of this figure is available).}
\end{figure*}

\renewcommand\thefigure{A\arabic{figure}}
\setcounter{figure}{0}

\begin{deluxetable*}{ccccc}
\tablenum{A1}
\tablecaption{Setup parameters of some PIC studies on plasma emission, including (from left to right) the number of cell (Nx$\times$Nz), the size of the domain (Lx$\times$Lz), the total simulation time, and NPPCPS for the background electrons and protons and the beam electrons.}
\tablewidth{0pt}
\tablehead{
\colhead{}  & \colhead{Nx$\times$Nz} & \colhead{Lx$\times$Lz ($\lambda_\mathrm{De}$)} & \colhead{Duration ($\omega_\mathrm{pe}^{-1}$)} & \colhead{NPPCPS}
}
\startdata
\citet{2001JGR...10618693K}   & 512$\times$512 & 512$\times$512 & $\sim$ 328 & 16, 16, 4\\
\citet{2009ApJ...694..618R}     & 512$\times$512 & 512$\times$512 & $\sim$ 328 & 80, 80, 8\\
\citet{2010JGRA..115.1204U}           & 1024$\times$1024 & 1024$\times$1024 & NA & 256, 256, 256\\
TT15                   & 600$\times$600   & 600$\times$600 & $\sim$1000 & 1000, 1000, 1000\\
\citet{2019JGRA..124.1475H}    & 1024$\times$1024 & 3072$\times$3072 &$\sim$1500 & 3600, 900, 900 \\
\citet{2022ApJ...924L..34C}     & 2048$\times$2048 & 6667$\times$6667 & 2000 & 2000, 1000, 1000\\
This paper             & 2048$\times$2048 & 6000$\times$6000 & 2000 & 4000, 1000, 1000\\
\enddata
\end{deluxetable*}

\begin{figure*}[ht]
 \centering
 \includegraphics[width=0.7\linewidth]{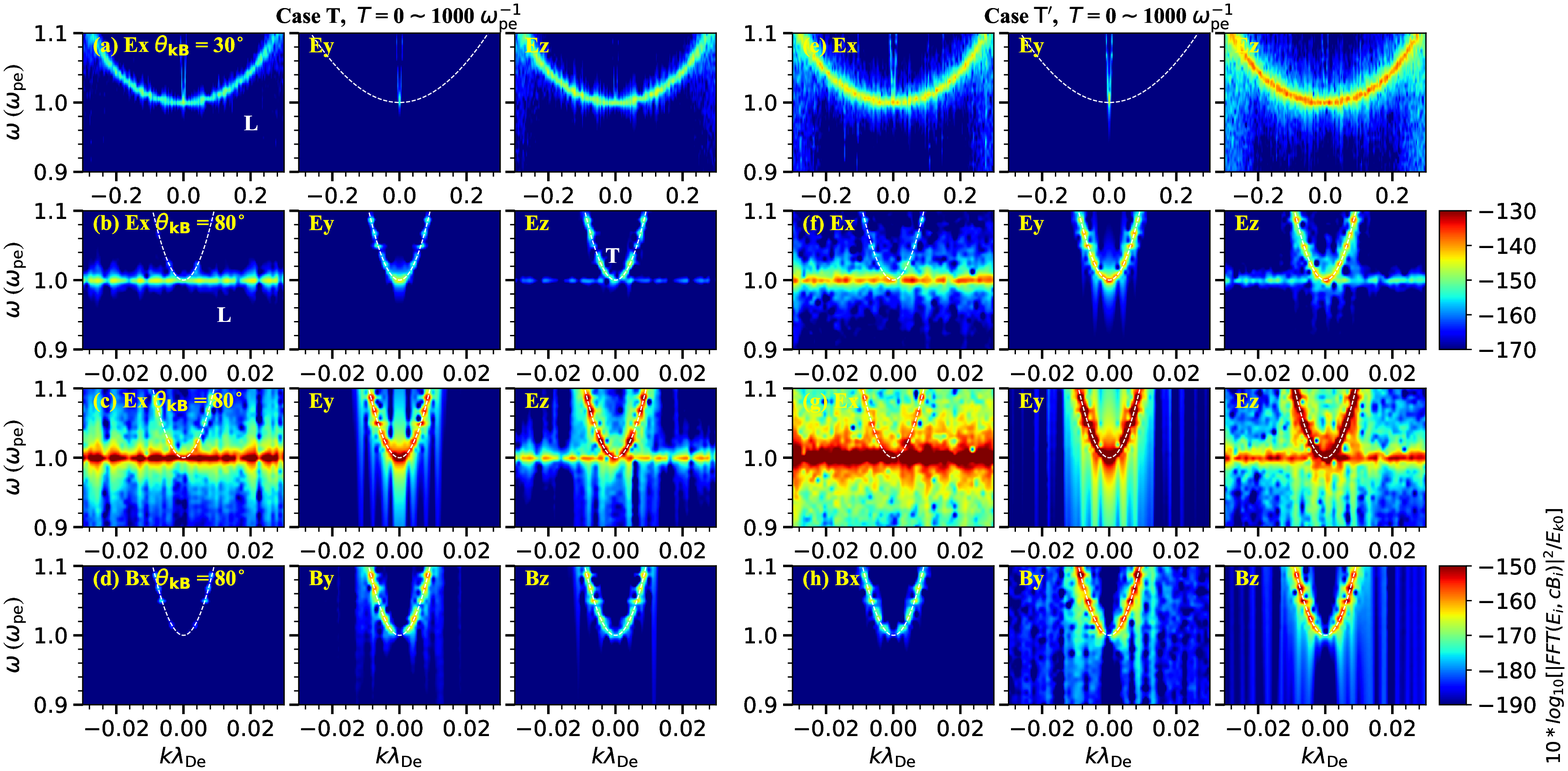}
  \caption{The same as Figure 4 yet for Cases T and T$^\prime$. The two cases are done with different sets of NPPCPS. In Case T the NPPCPS = 4000 (for electrons) and 1000 (for protons); in Case T$^\prime$ the NPPCPS = 500 (for electrons) and 125 (for protons). The video begins at $\theta = 0^\circ$ and advances $5^\circ$ at a time up to $\theta = 90^\circ$. The real-time duration of the video is 5 s. (An animation of this figure is available).}
\end{figure*}

\begin{figure*}[ht]
 \centering
 \includegraphics[width=0.7\linewidth]{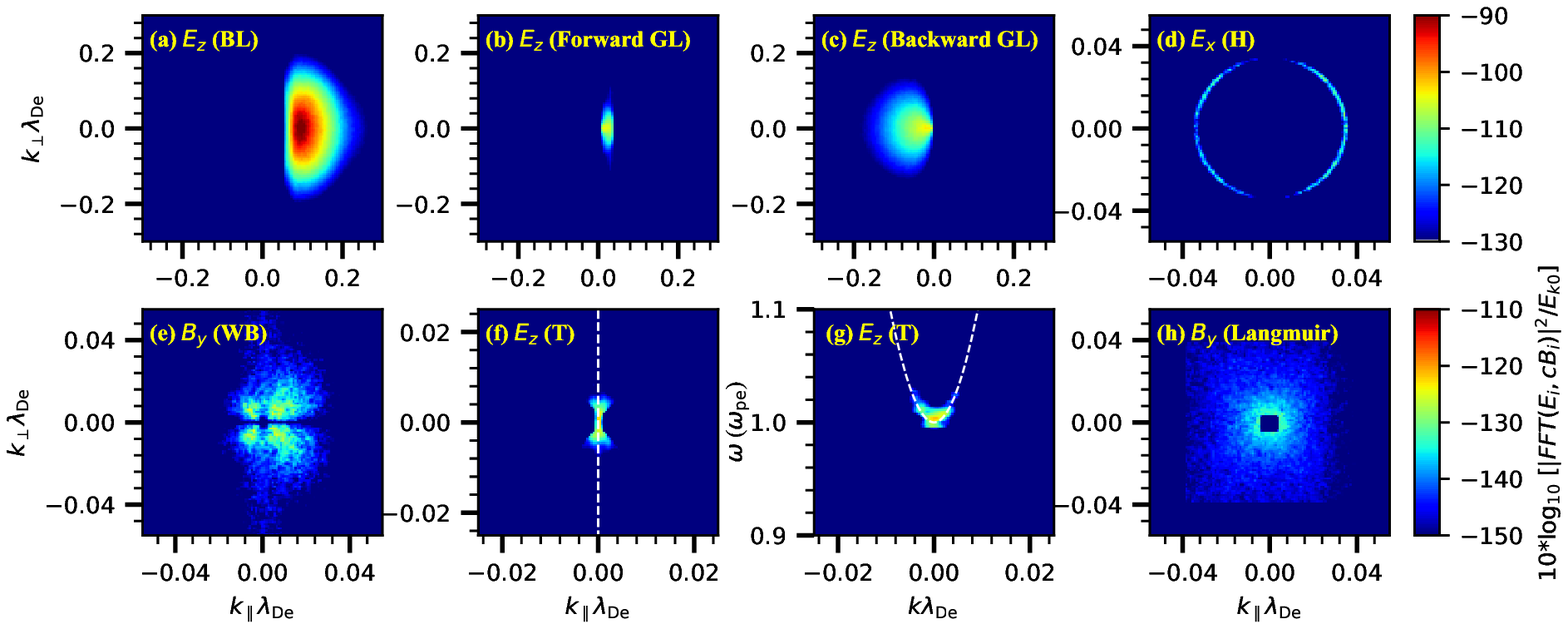}
  \caption{The spectral regime in the $\vec k$ space selected to evaluate the energy of various wave modes. The spectral region indicated by the black square in the last panel is excluded when calculating the mode energy.}
\end{figure*}

\begin{figure*}[ht]
 \centering
 \includegraphics[width=0.7\linewidth]{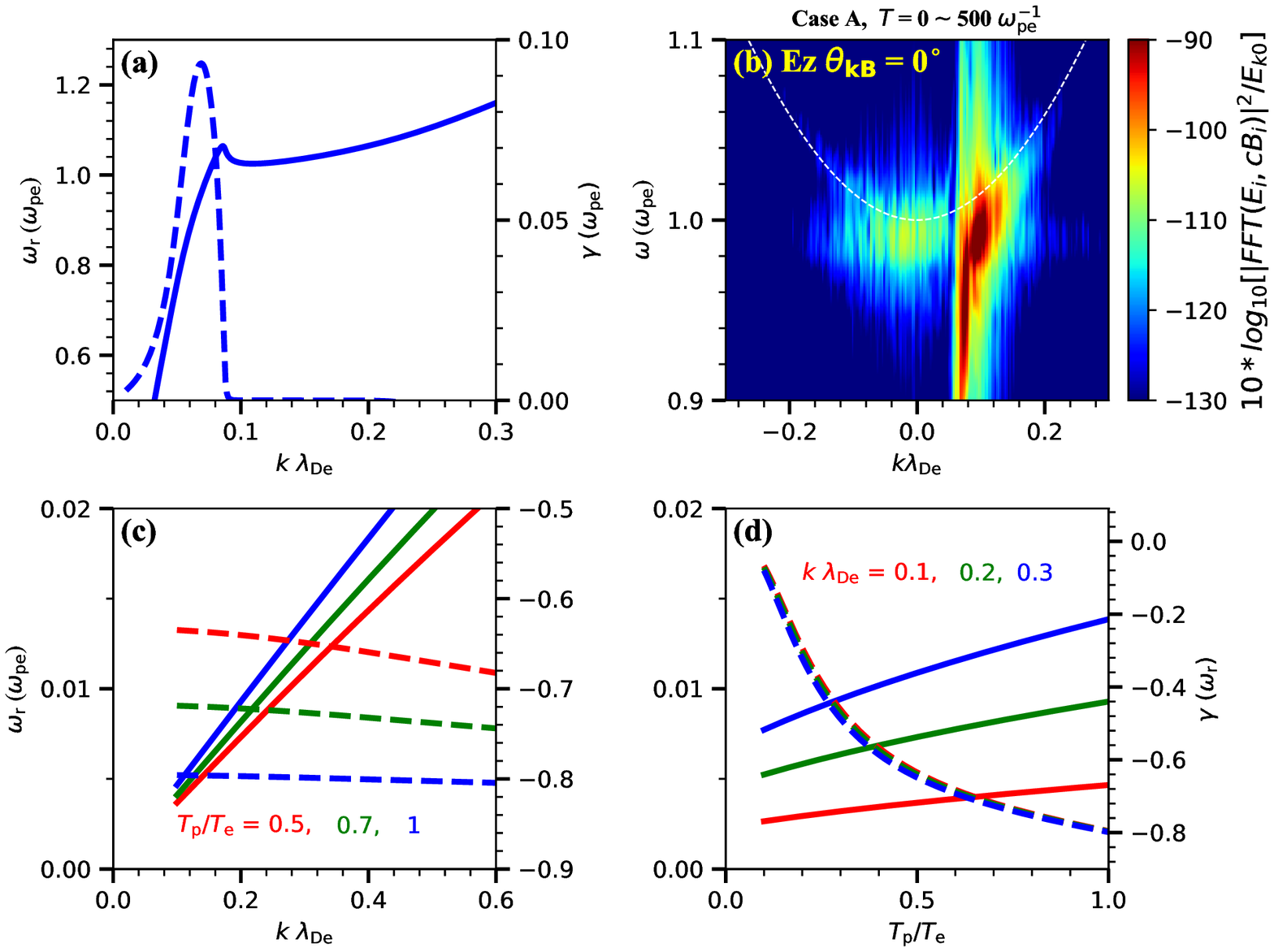}
  \caption{(a) the growth rate ($\gamma$: dashed line) and frequency ($\omega_\mathrm{r}$: solid line) of the beam-Langmuir mode excited via the kinetic bump-on-tail instability; (b) the dispersion diagram of Case A within the period of [0, 500 $\omega_\mathrm{pe}^{-1}$]); (c) and (d), the growth rate ($\gamma$: dashed line) and frequency ($\omega_\mathrm{r}$: solid line) of the IA mode, versus the normalized wave number (c) and the proton-electron temperature ratio (d). }
\end{figure*}

\begin{figure*}[ht]
 \centering
 \includegraphics[width=0.7\linewidth]{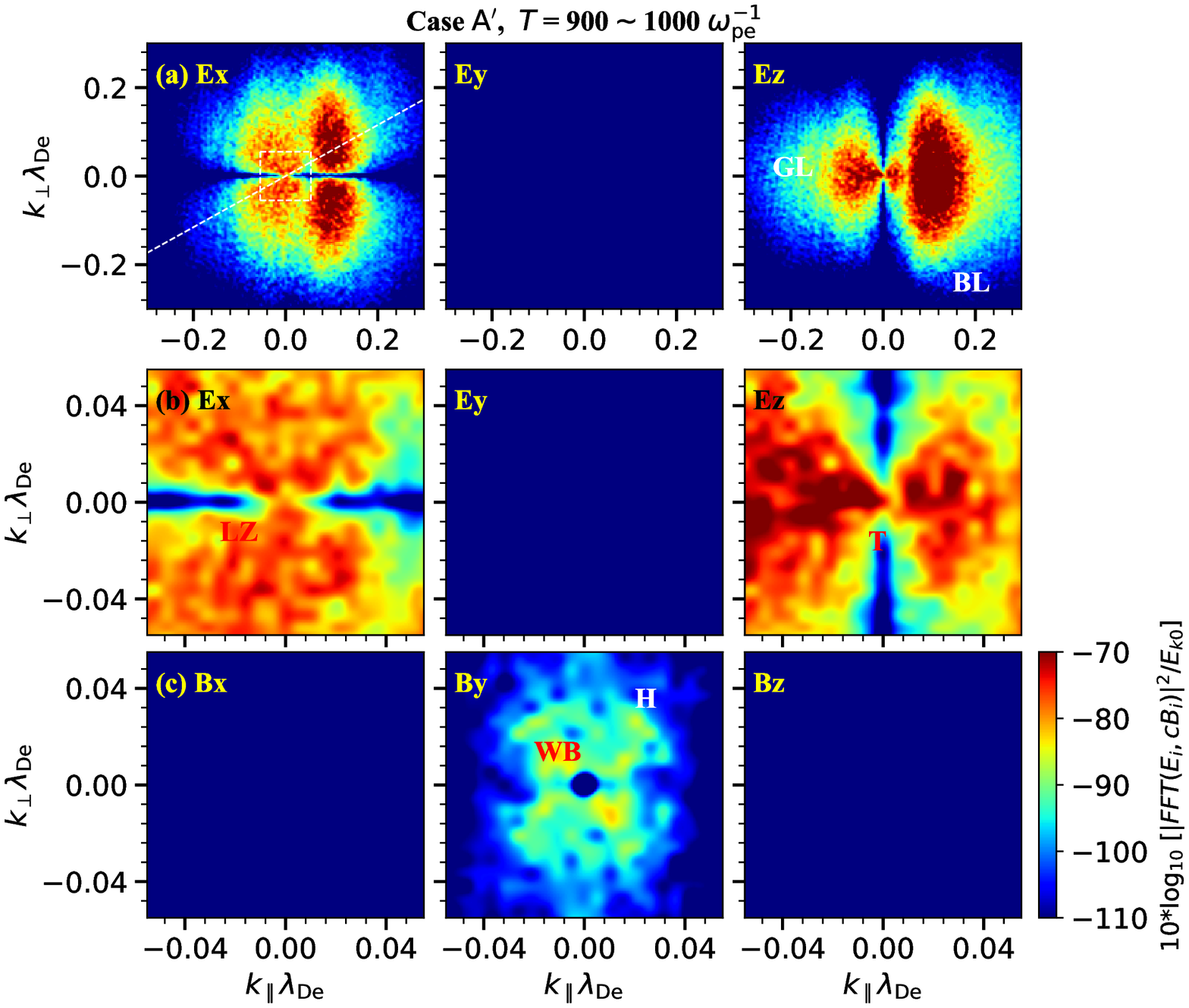}
  \caption{The same as Figure 3 yet for Case A$^\prime$.}
\end{figure*}

\begin{figure*}[ht]
 \centering
 \includegraphics[width=0.7\linewidth]{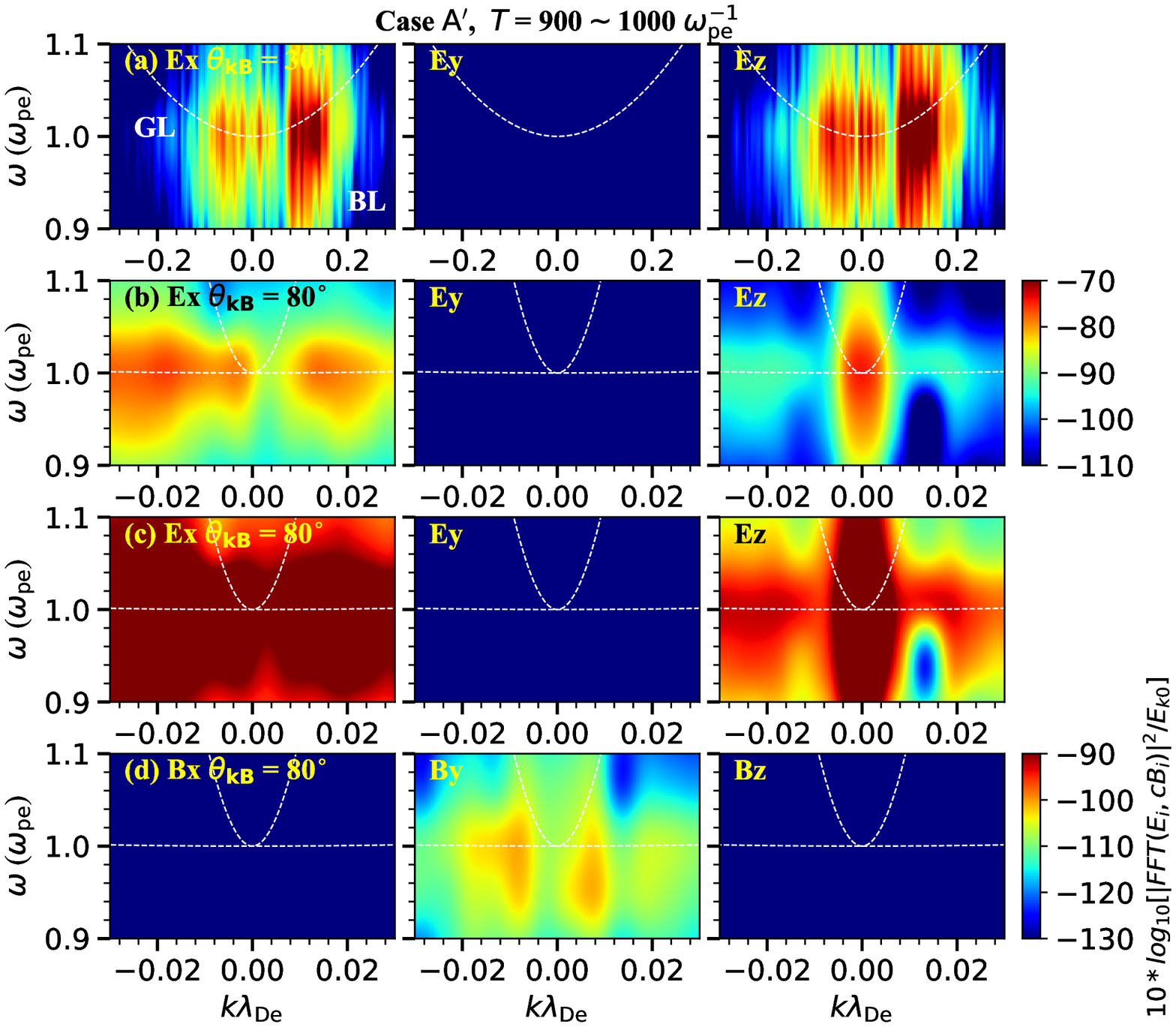}
  \caption{The same as Figure 4 yet for Case A$^\prime$. The video begins at $\theta = 0^\circ$ and advances $5^\circ$ at a time up to $\theta = 90^\circ$. The real-time duration of the video is 5 s. (An animation of this figure is available).}
\end{figure*}

\end{document}